\documentclass[a4paper]{spie}
\usepackage[]{graphicx}

\title{Speckle nulling for exoplanet detection \\ with space-based coronagraphic telescopes} 

\author{Pascal J. Bord\'e\supit{a,b}, Wesley A. Traub\supit{a,c}, Brian D. Kern\supit{c}, \\ John T. Trauger\supit{c}, and Andreas C. Kuhnert\supit{c}
\skiplinehalf
\supit{a}Harvard-Smithsonian Center for Astrophysics, 60 Garden Street, Cambridge, MA 02138 \\
\supit{b}California Institute of Technology, MS 100--22, 770 S Wilson Avenue, Pasadena, CA 91125 \\
\supit{c}Jet Propulsion Laboratory, MS 301--451, 4800 Oak Grove Drive, Pasadena, CA 91109
}

\authorinfo{Further author information: (Send correspondence to P.J.B.)\\P.J.B.: Michelson Postdoctoral Fellow, E-mail: pborde@cfa.harvard.edu \\
Copyright 2006 Society of Photo-Optical Instrumentation Engineers. \\
This paper will be published in the proceedings of the conference
\emph{Space Telescopes and Instrumentation I: Optical, Infrared, and Millimeter}, part of SPIE's Astronomical
Telescopes \& Instrumentation, 24--31 May 2006, Orlando, FL and is made
available as an electronic preprint with permission of SPIE. One print or
electronic copy may be made for personal use only. Systematic or multiple
reproduction, distribution to multiple locations via electronic or other means,
duplication of any material in this paper for a fee or for commercial purposes,
or modification of the content of the paper are prohibited.
}

\begin{document} 
  \maketitle 

\begin{abstract}
To detect Earth-like planets in the visible with a coronagraphic telescope, two major noise sources have to be overcome: the photon noise of the diffracted star light, and the speckle noise due to the star light scattered by instrumental defects. Coronagraphs tackle only the photon noise contribution. In order to decrease the speckle noise below the planet level, an active control of the wave front is required. We have developed analytical methods to measure and correct the speckle noise behind a coronagraph with a deformable mirror. In this paper, we summarize these methods, present numerical simulations, and discuss preliminary experimental results obtained with the High-Contrast Imaging Testbed at NASA's Jet Propulsion Laboratory.  
\end{abstract}

\keywords{High-contrast imaging, Coronagraphy, Adaptive Optics, Exoplanets}

\section{INTRODUCTION}
\label{sec:intro}

In coronagraphs like any other optical system, aberrations give rise to scattered light in the form of speckles in the focal plane, thus preventing very high-contrast imaging. A wave front sensing and control system can remedy this, and make the detection of faint objects possible\cite{Malbet95}. Such a system is absolutely critical to the success of NASA's Terrestrial Planet Finder Coronagraph (TPF-C)\cite{Coulter04} that aims not only at the detection, but also at the spectroscopy of terrestrial planets in the visible range. This is a very challenging task as the contrast\footnote{The contrast is defined here as ratio of the luminosity of the star to that of the planet at the wavelength of interest.} between a terrestrial planet and its Sun-like star amounts to $\sim 10^{10}$ in the visible. With that goal in mind, we have devised a theory\cite{Borde06} to measure and correct wave front aberrations, so as to clear out of speckles an area of the focal plane, referred to as dark hole (DH). This area is then suitable for very faint object detection.

In the following, we first summarized the theory (\S\ref{sec:theory}), then we present simulations (\S\ref{sec:simulations}), and finally we discuss preliminary experimental results (\S\ref{sec:experiments}).

\section{THEORY SUMMARY} 
\label{sec:theory}

The wave front sensing and control system we advocate consists of
\begin{enumerate}
\item The sensor: the science camera itself since a separate sensing channel would have its own aberrations, impossible to disentangle from the main optical train's aberrations (\S\ref{sub:measurement});
\item The actuator: a deformable mirror (DM) with $N\!\times\!N$ individual actuators (\S\ref{sub:dm});
\item The command control system: a computer converting speckle field measurements into DM commands (\S\ref{sub:nulling}).
\end{enumerate}
Let us now examine briefly these three elements.

\subsection{Speckle field measurement}
\label{sub:measurement}

In a regime of small aberrations, the electric field in the image plane after the coronagraph is the sum of two terms: the original speckle field, $\hat{\Phi}$, and the DM-controlled electric field, $\hat{\Psi}$. In principle, three images with different DM settings bring enough information to measure the complex amplitude of the speckle field. Indeed, the intensity $I_n$ ($n=0,1,2$) in any given pixel of image $n$ reads
\begin{equation} \label{eq:system_I}
\left \{
\begin{array}{l}
I_0 = |\hat{\Phi} + \hat{\Psi}_0|^2 \\
I_1 = |\hat{\Phi} + \hat{\Psi}_0 + \delta\hat{\Psi}_1|^2 \\
I_2 = |\hat{\Phi} + \hat{\Psi}_0 + \delta\hat{\Psi}_2|^2. \\
\end{array}
\right.
\end{equation}
System (\ref{eq:system_I}) has for solution
\begin{equation}
\hat{\Phi} = \frac{\delta\hat{\Psi}_2 \, (I_1 - I_0 - |\delta\hat{\Psi}_1|^2) -
\delta\hat{\Psi}_1 \, (I_2 - I_0 - |\delta\hat{\Psi}_2|^2)}{\delta\hat{\Psi}_1^\ast \delta\hat{\Psi}_2 - \delta\hat{\Psi}_1 \delta\hat{\Psi}_2^\ast} - \hat{\Psi}_0,
\end{equation}
with the caveat that the denominator should not be zero. This tells us that the successive modifications in the DM-controlled electric field, $\delta\hat{\Psi}_1$ and $\delta\hat{\Psi}_2$, must lead to an intensity change in any given pixel big enough so that information could be acquired for that pixel during the measurement process.

\subsection{Deformable mirror model}
\label{sub:dm}

Assuming linear superposition and introducing actuator influence functions\footnote{An influence function describes the DM surface deformation in response to the actuation of a given actuator.}, the DM-controlled electric field can be expanded as
\begin{equation} \label{eq:psi}
\hat{\Psi} = \sum_{k=0}^{N-1} \sum_{l=0}^{N-1} a_{kl} \hat{F}_{kl},
\end{equation}
where $a_{kl}$ are the actuator strokes and $\hat{F}_{kl}$ the complex images on the detector of the influence functions. For the simulations and the experiments, we used a $32\!\times\!32$ DM with a 1-mm actuator spacing. More information about the DM can be found in Ref.~\citenum{Trauger03}.

\subsection{Speckle nulling}
\label{sub:nulling}

\paragraph{Field nulling} In this approach, we try to cancel out the speckle field in every detector pixel. The basic equation is simply $\hat{\Phi} + \hat{\Psi} = 0$, i.e. by making use of Eq.~(\ref{eq:psi}),
\begin{equation} \label{eq:field_nulling} 
\sum_{k=0}^{N-1} \sum_{l=0}^{N-1} a_{kl} \hat{F}_{kl} = -\hat{\Phi}. 
\end{equation}
Equation (\ref{eq:field_nulling}) is a linear system in the actuator strokes that can be solved either by singular value decomposition\cite{Press97}, or very efficiently with an inverse Fast Fourier Transform (FFT), provided $\hat{F}_{kl}$ are phase-shifted copies of one another. Although this last hypothesis does not rigorously hold behind a coronagraph, simulations show that the solution by inverse FFT could yield satisfactory results for a first round of tests.

\paragraph{Energy minimization} In this approach, we seek to minimize the total energy of the speckles in the DH. This energy is defined as
\begin{equation} \label{eq:energy1}
\mathcal{E} \equiv \int\!\!\!\int_\mathrm{DH} |\hat{\Phi} + \hat{\Psi}|^2.
\end{equation}
Equation (\ref{eq:energy1}) is minimized when the energy gradient with respect to the actuator strokes is zero, i.e.
\begin{equation} \label{eq:energy2}
\forall (k,l) \in {\{0 \ldots N\!-\!1\}}^2, \quad \frac{\partial \mathcal{E}}{\partial a_{kl}} = 0
\quad \Longleftrightarrow \quad \sum_{n=0}^{N-1} \sum_{m=0}^{N-1} a_{nm} \int\!\!\!\int_\mathrm{DH} \hat{F}_{nm} \hat{F}_{kl}^\ast =
-\mbox{Re} \left( \int\!\!\!\int_\mathrm{DH} \hat{\Phi} \hat{F}_{kl}^\ast \right).
\end{equation}
Equation (\ref{eq:energy2}) is again a linear system in the actuator strokes, but with smaller matrices than in equation (\ref{eq:field_nulling}) because matrix sizes are governed here solely by the number of actuators, and no longer by the number of detector pixels. Simulations show that speckle energy minimization is more powerful and flexible than speckle field nulling. We intend to demonstrate experimentally this second approach in the near future.

\section{SIMULATIONS}
\label{sec:simulations}

\begin{figure}[p]
\begin{center}
\begin{tabular}{c}
\includegraphics[width=4cm]{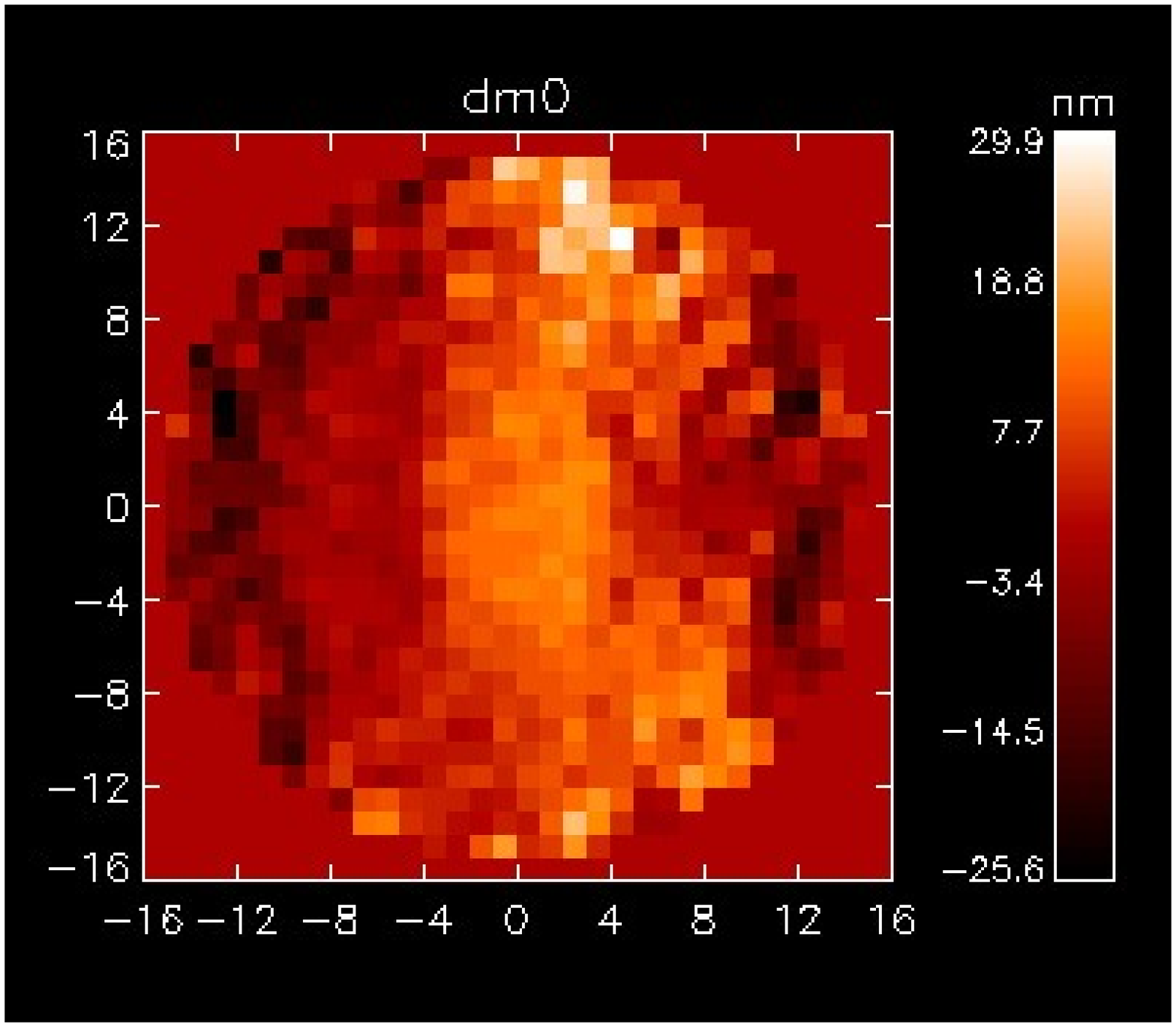}
\includegraphics[width=4cm]{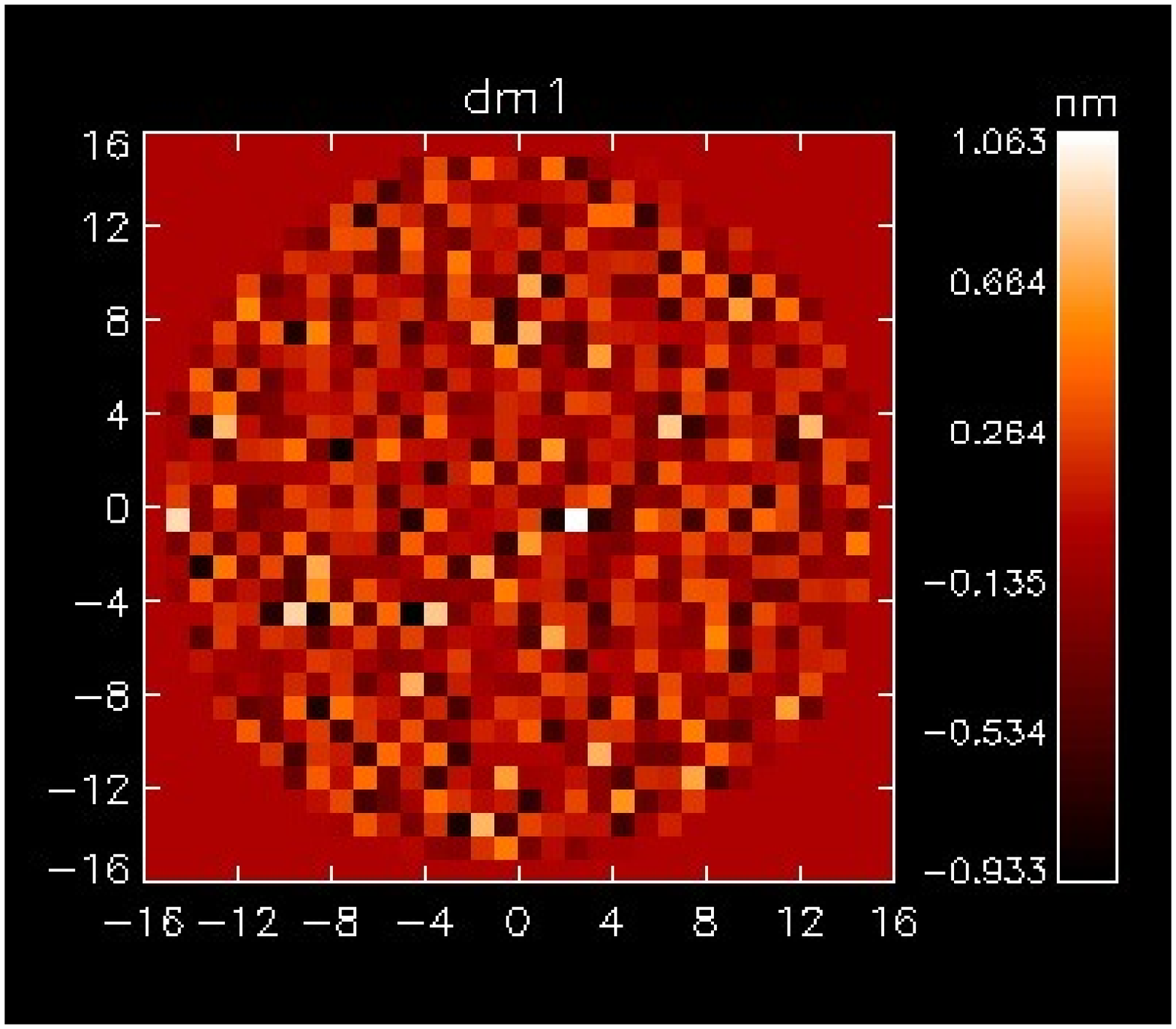}
\includegraphics[width=4cm]{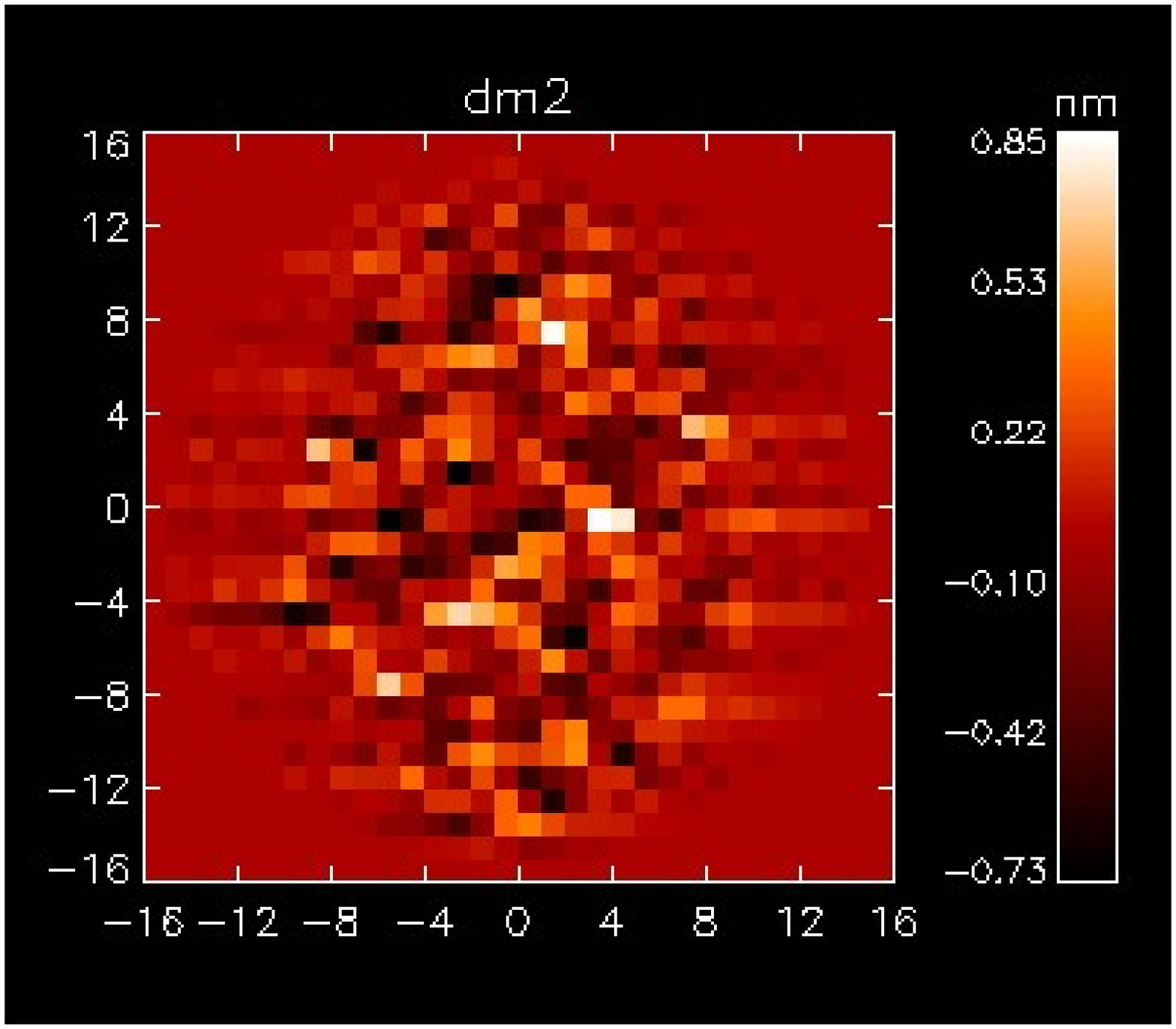}
\includegraphics[width=4cm]{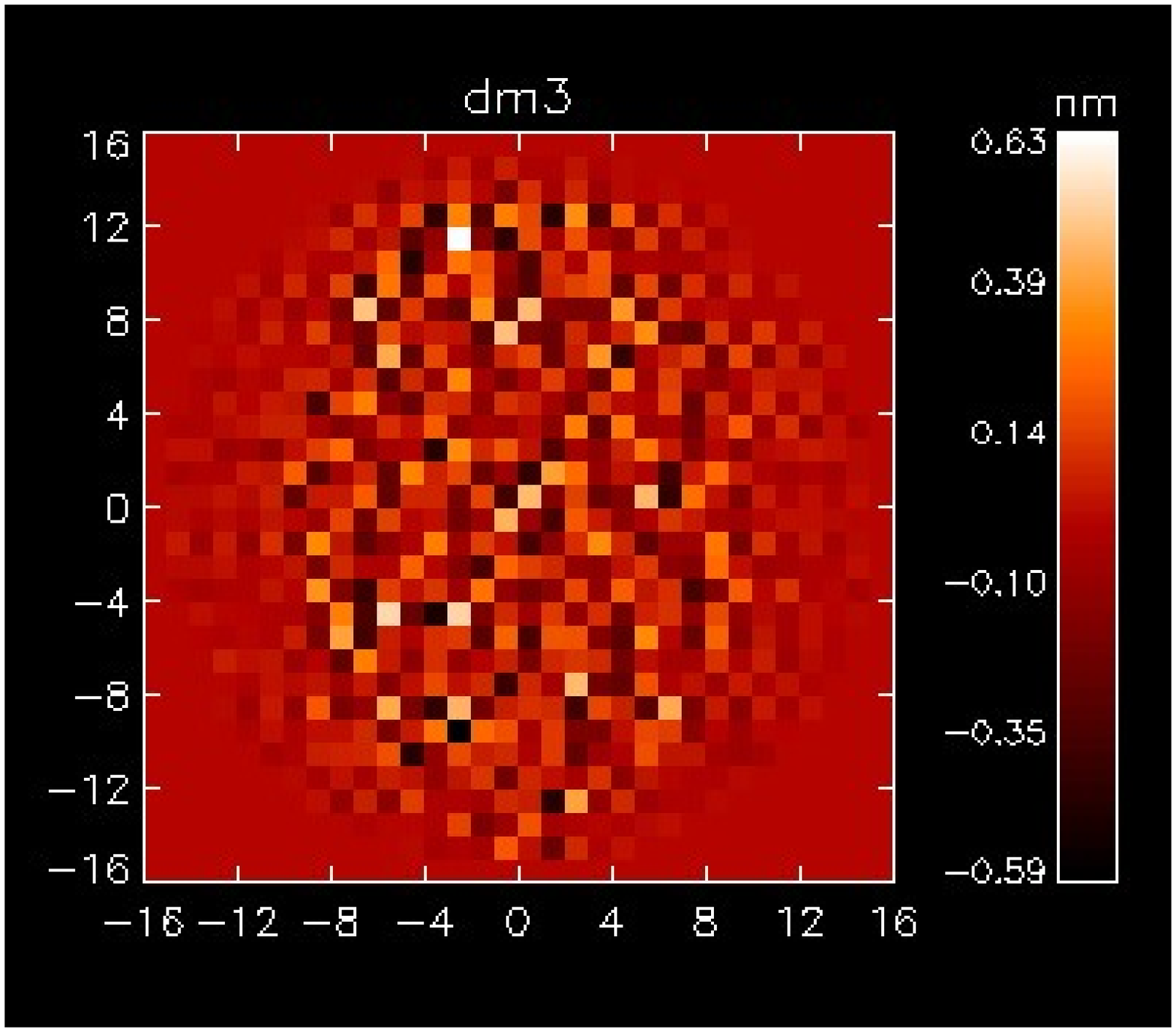}
\end{tabular}
\end{center}
\caption[example] { \label{fig:sim_dm} 
Simulated series of deformable mirror (DM) settings during the speckle nulling process. The first image on the left displays the initial DM setting (dm0), whereas other images (dm1, dm2, \& dm3) show changes with respect to that initial setting. Note that the DM is not initially flat because the bulk of low-order aberrations has already been taken out. In the second and third steps (dm1 \& dm2), the speckle field is modified so as to measure its complex amplitude. In the fourth step (dm3), the shape imposed on the DM creates the dark hole. The DM features $32\!\times\!32$ actuators, but actuators outside the entrance pupil (a 30-mm diaphragm on top of the DM) are not commanded. Actuator strokes are color-coded in nm.}
\end{figure}

\begin{figure}[p]
\begin{center}
\begin{tabular}{cc}
\includegraphics[width=6cm]{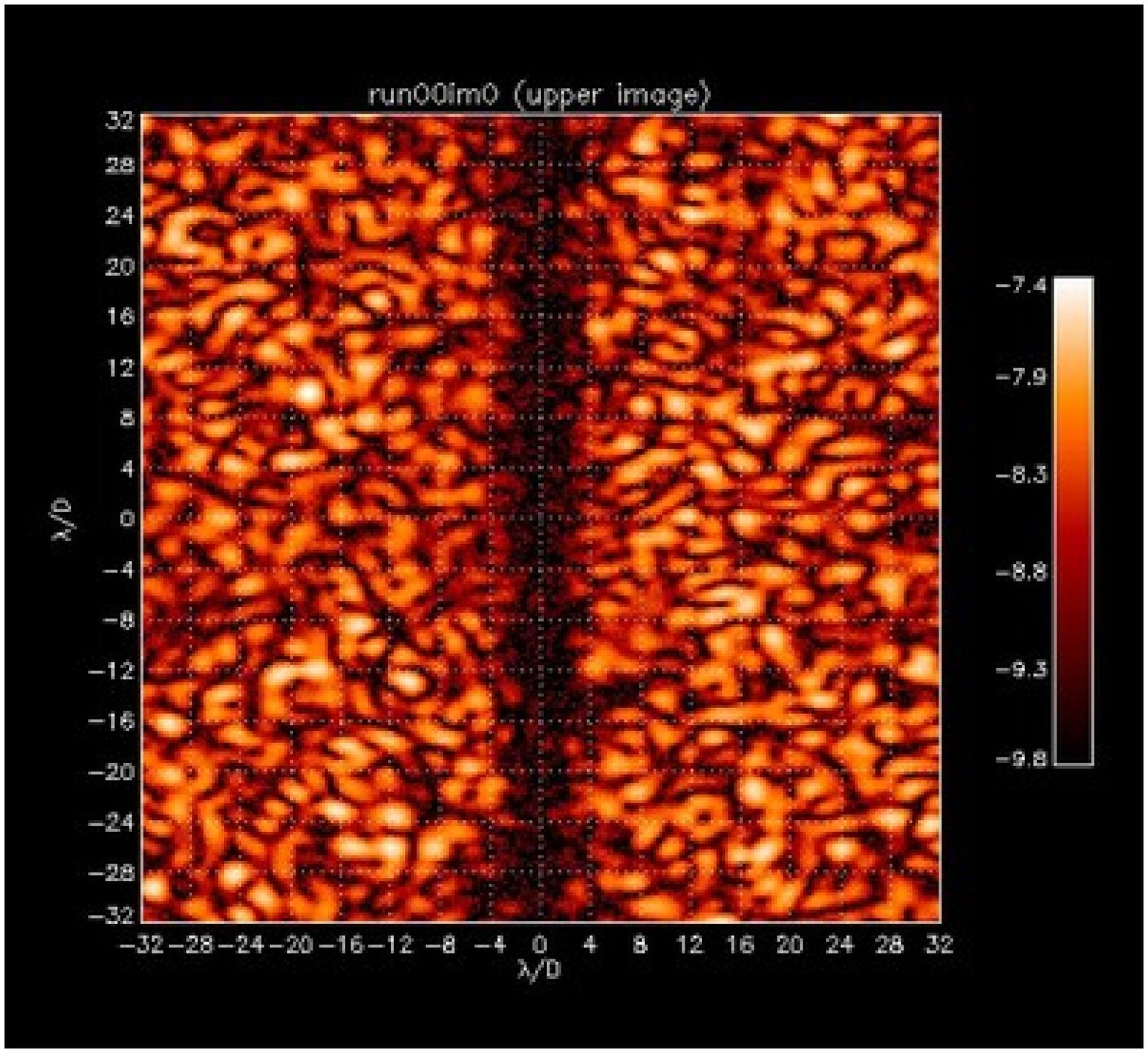} & \includegraphics[width=6cm]{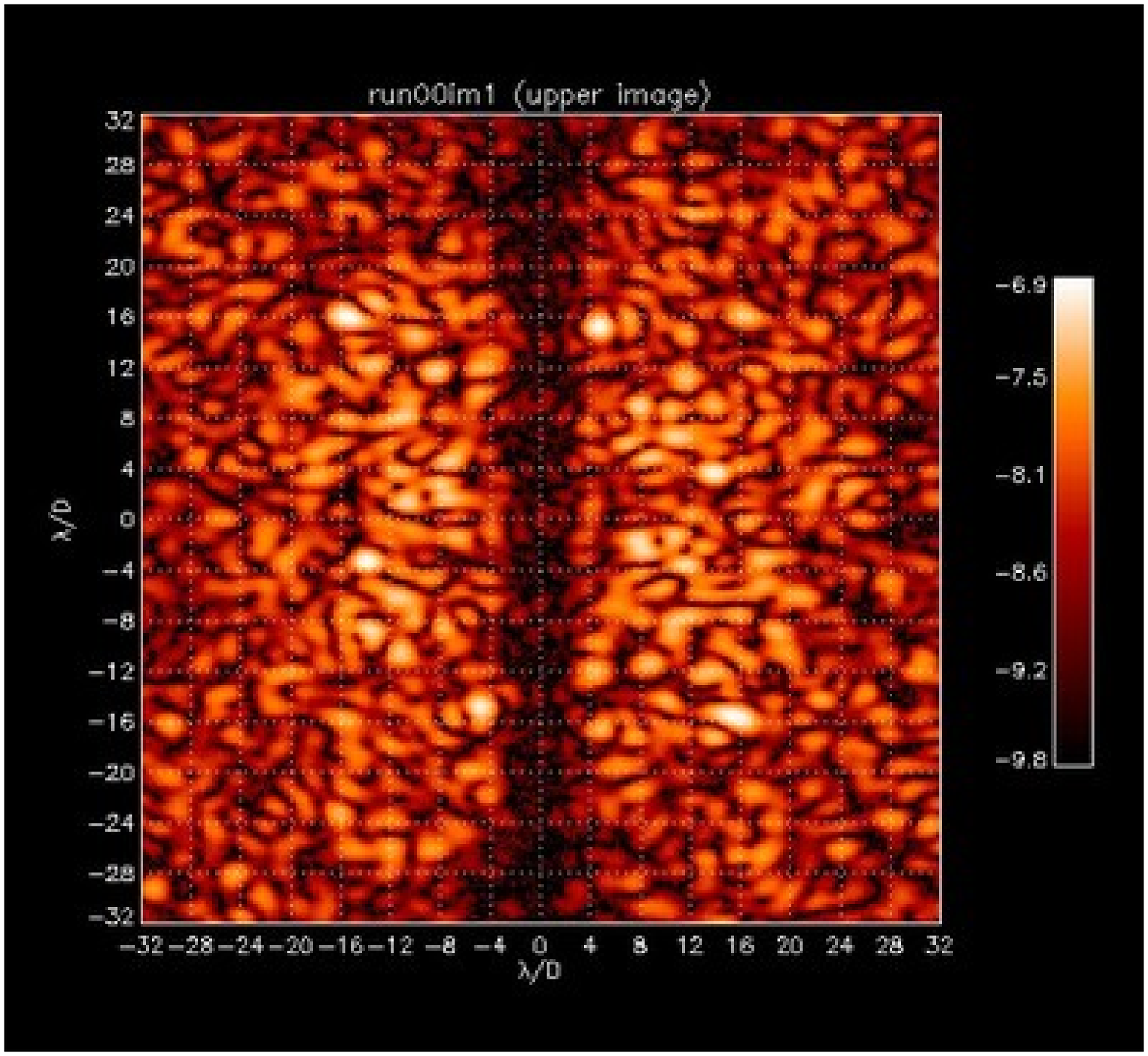} \\
\includegraphics[width=6cm]{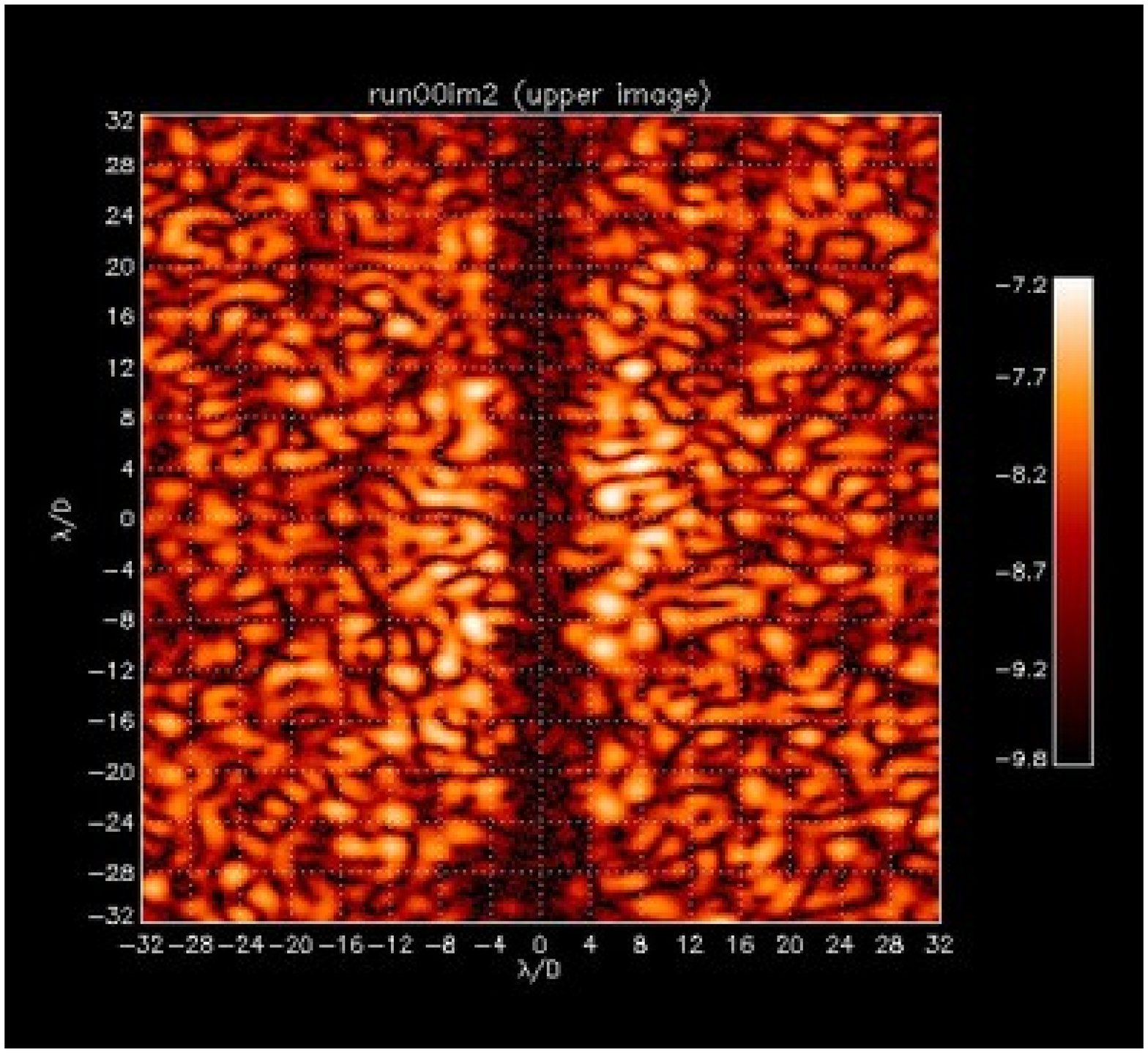} & \includegraphics[width=6cm]{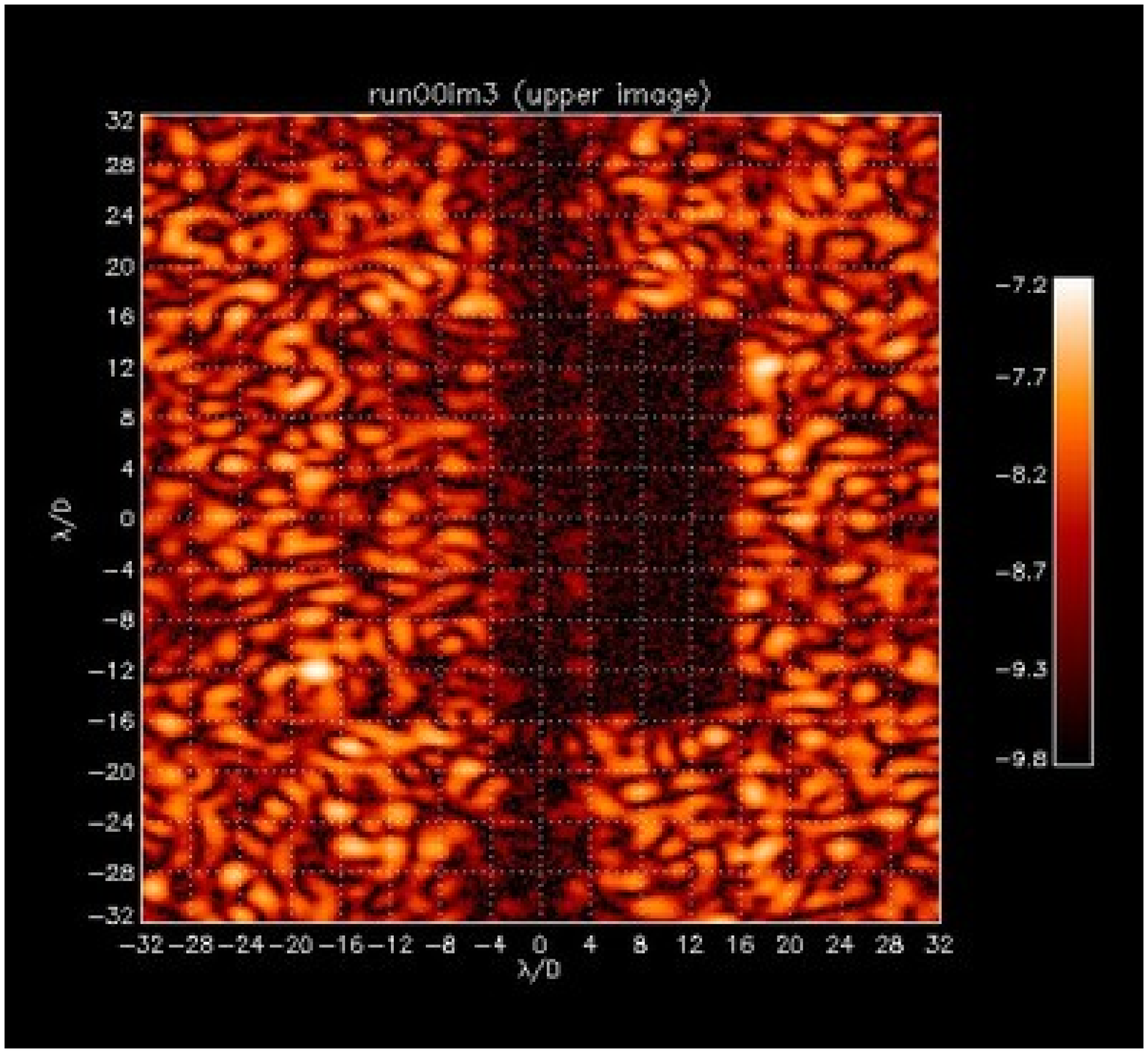}
\end{tabular}
\end{center}
\caption[example] { \label{fig:sim_im} 
Simulated series of detector images during the speckle nulling process. Every image displays the log of the contrast as defined in text. Superimposed is a dotted grid scaled in elements of resolution from $-32\frac{\lambda}{D}$ to $+32\frac{\lambda}{D}$. The top-left image shows the initial state: the speckle field covers the whole image except a vertical streak where the coronagraphic mask is opaque. In the second (top-right) and third (bottom-left) images, the speckle field is modified so as to measure its complex amplitude. In the fourth (bottom-right) image, the dark hole is apparent. Only the right side is corrected for both amplitude and phase aberrations.}
\end{figure}

In this section, we present monochromatic simulations using the field nulling equation (Eq.~\ref{eq:field_nulling}) solved with an inverse FFT (Figs.~\ref{fig:sim_dm}--\ref{fig:sim_im}). We consider a Lyot-type coronagraph with a band-limited image-plane mask\cite{Kuchner02}. Because the DM features $32\!\times\!32$ actuators on a square grid, the sampling theorem imposes that the maximum size for the DH is a $32\frac{\lambda}{D} \times 32 \frac{\lambda}{D}$ square, where $\lambda$ is the wavelength and $D$ the entrance pupil's diameter. When both phase and amplitude defects are present, a single DM creates a DH limited to half the maximum size. The simulations in Figs.~\ref{fig:sim_dm}--\ref{fig:sim_im} are meant to illustrate the theory, not to be realistic. They assume a Gaussian white noise for phase defects and amplitude defects, but no photon nor detector noise. The contrast in any given pixel is defined as the ratio of the intensity in that particular pixel to the peak intensity measured in the image when the coronagraph image-plane mask is removed. In the next section, we discuss preliminary experimental results and compare them with more realistic simulations.

\section{EXPERIMENTS}
\label{sec:experiments} 

\begin{figure}[p]
\begin{center}
\begin{tabular}{c}
\includegraphics[width=4cm]{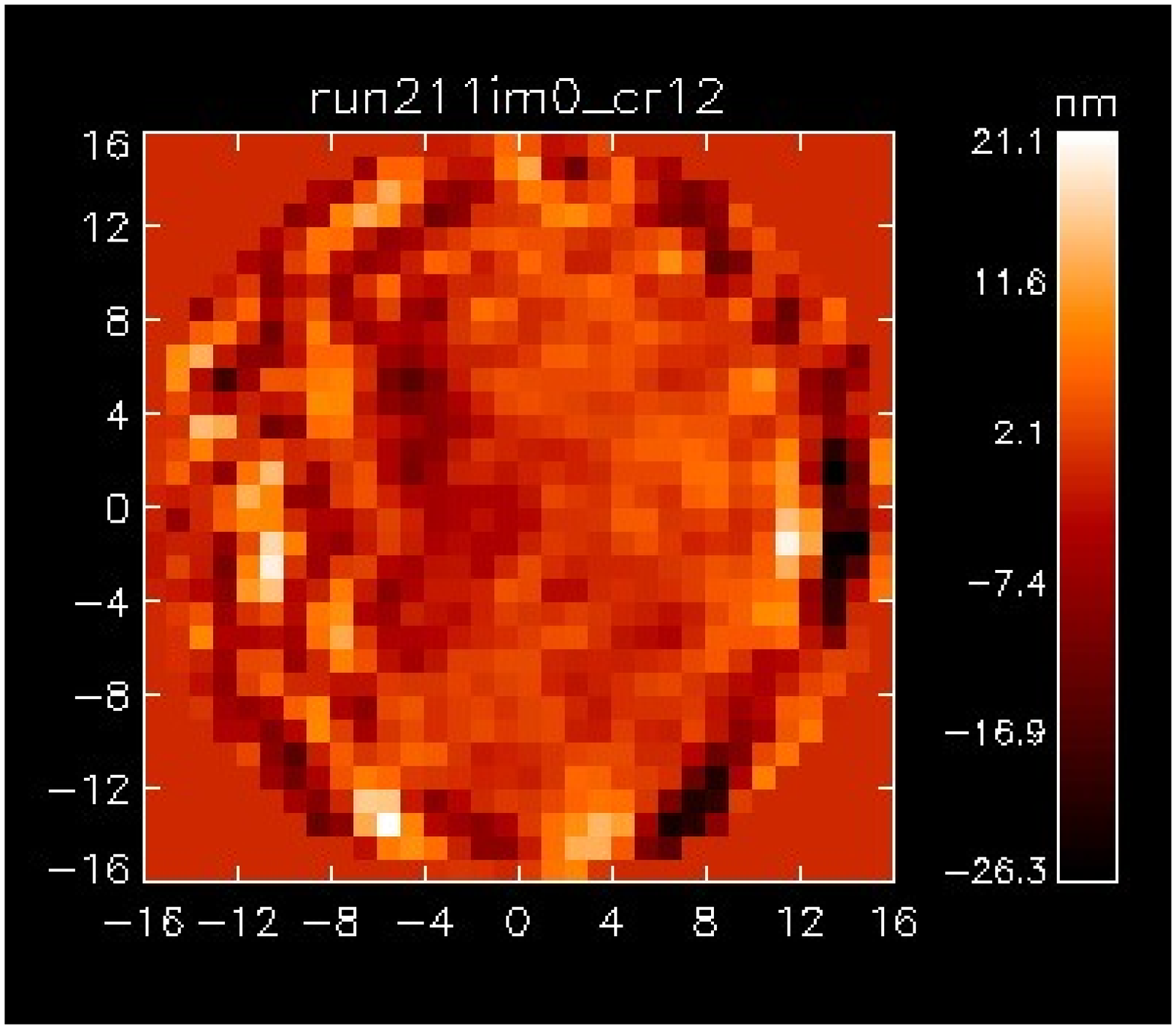}
\includegraphics[width=4cm]{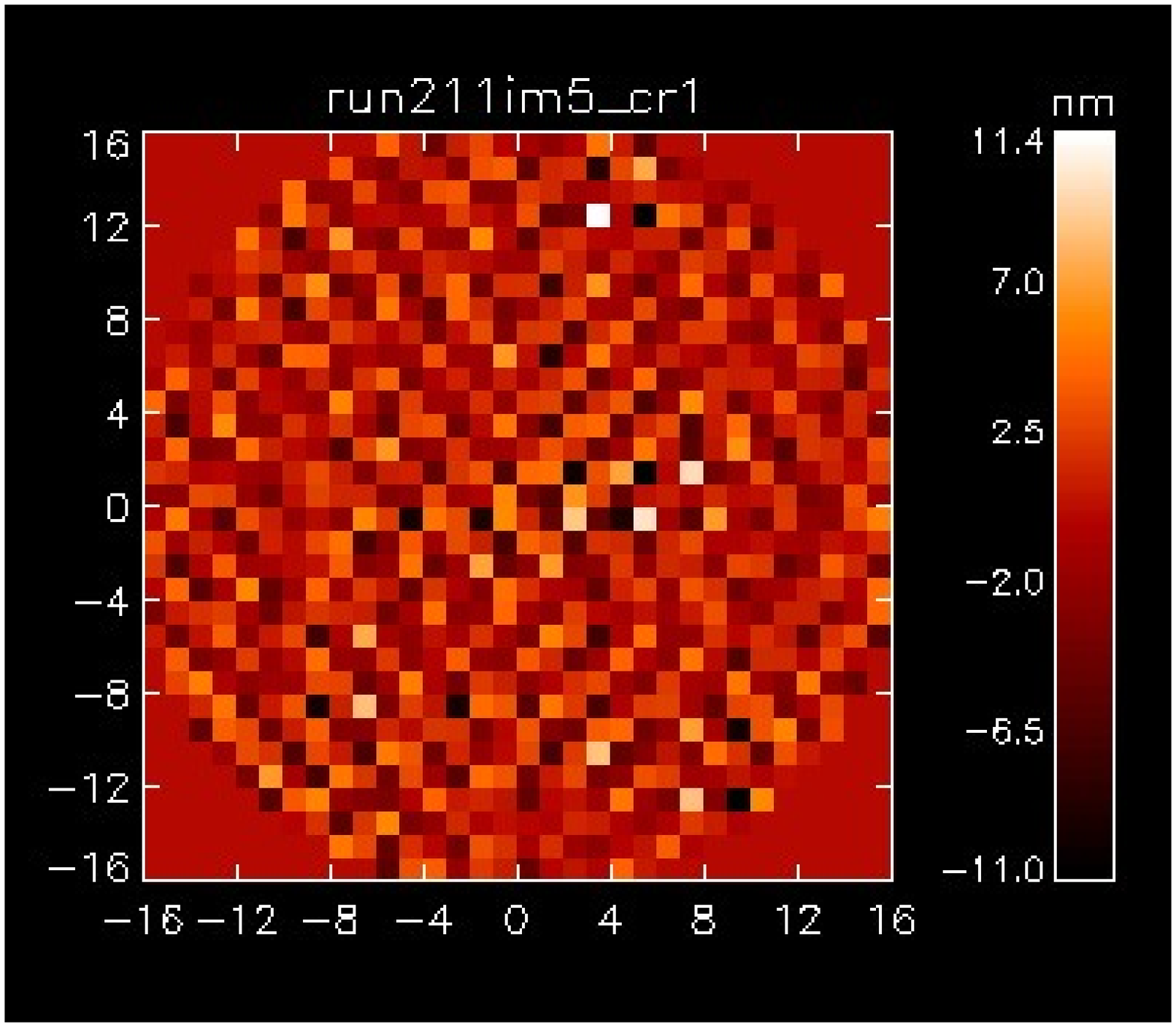}
\includegraphics[width=4cm]{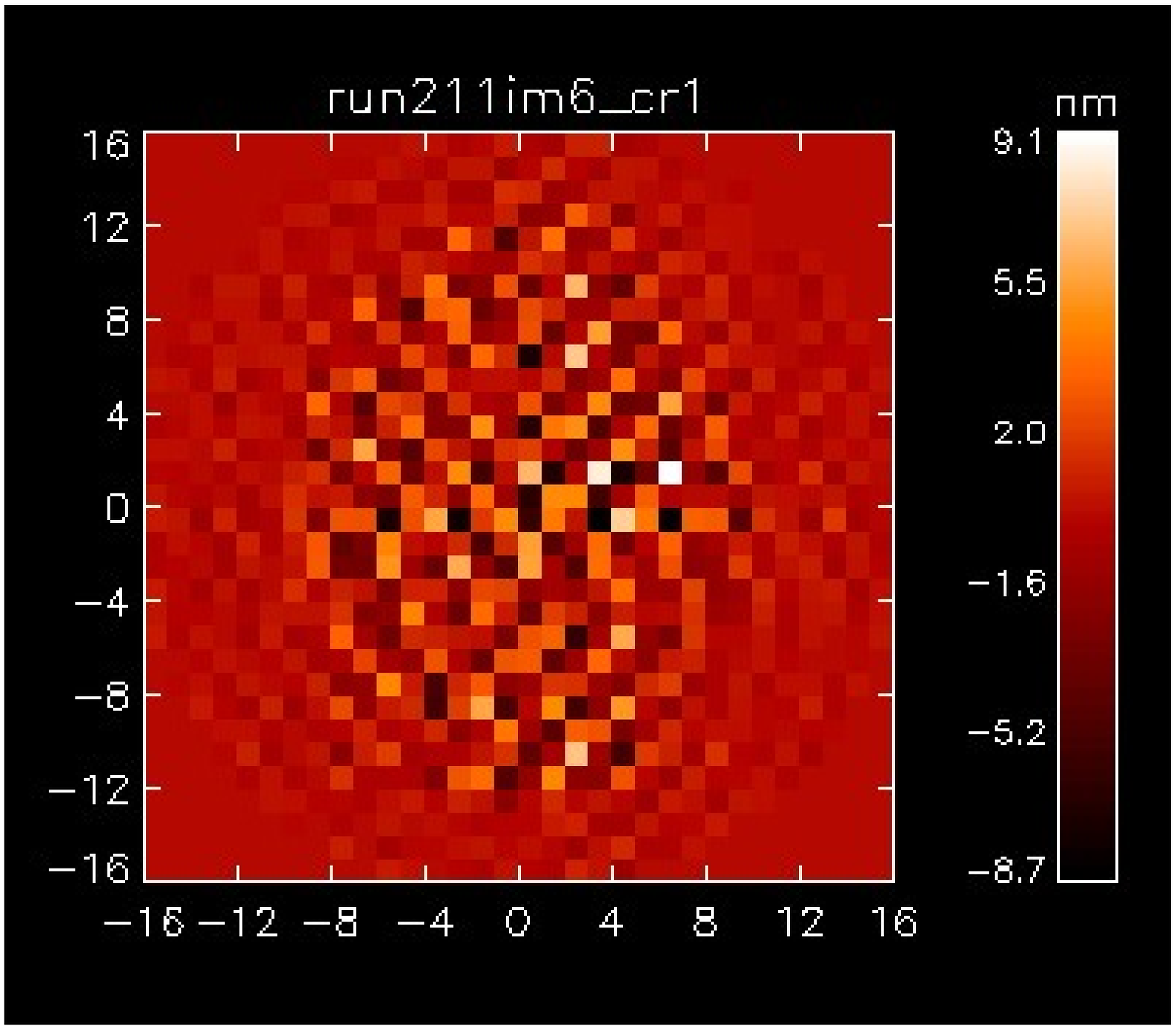}
\includegraphics[width=4cm]{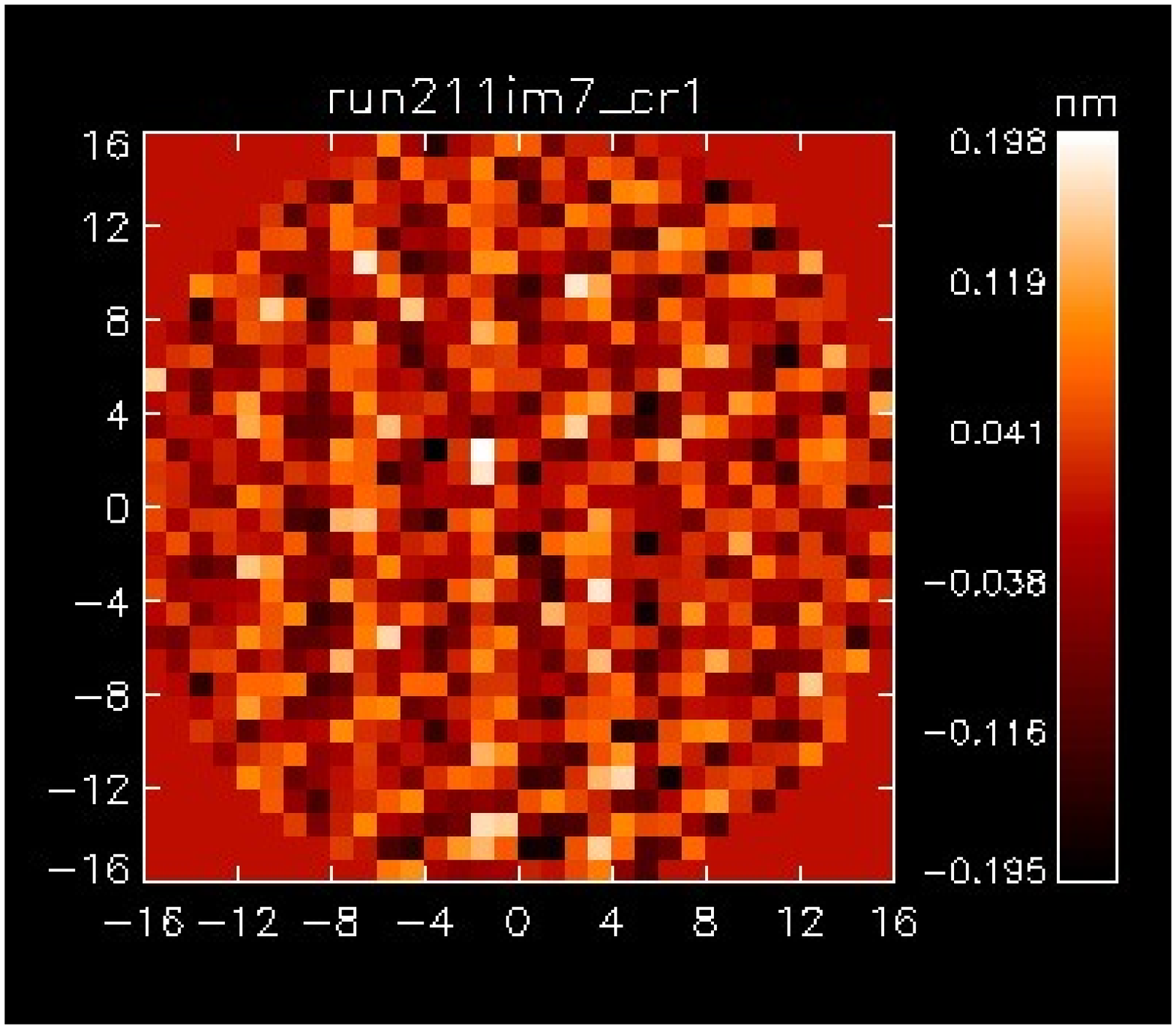}
\end{tabular}
\end{center}
\caption[example] { \label{fig:exp_dm} 
Same as Fig.~\ref{fig:sim_dm} but for a real experiment.}
\end{figure}

\begin{figure}[p]
\begin{center}
\begin{tabular}{cc}
\includegraphics[width=6cm]{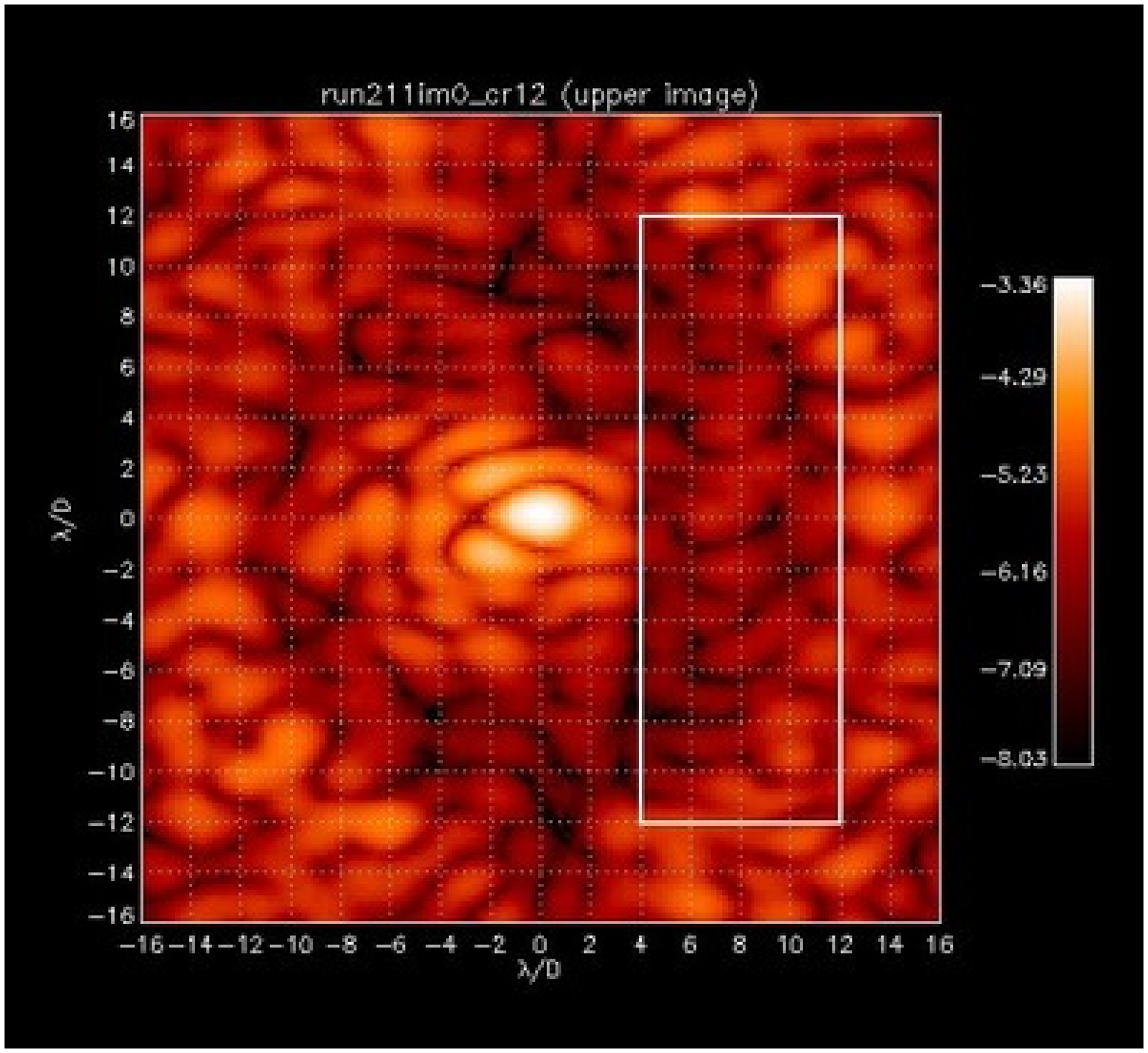} & \includegraphics[width=6cm]{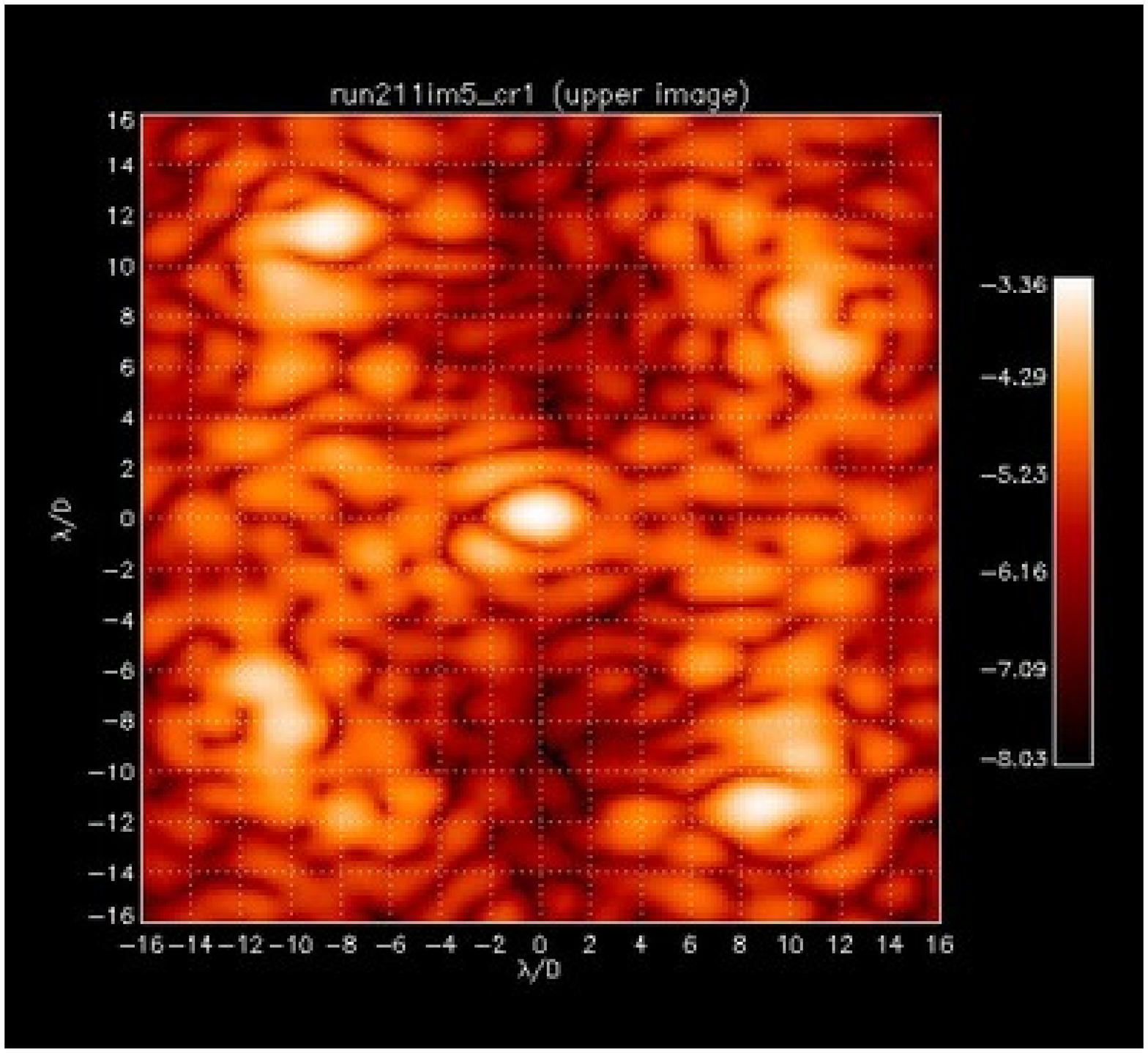} \\
\includegraphics[width=6cm]{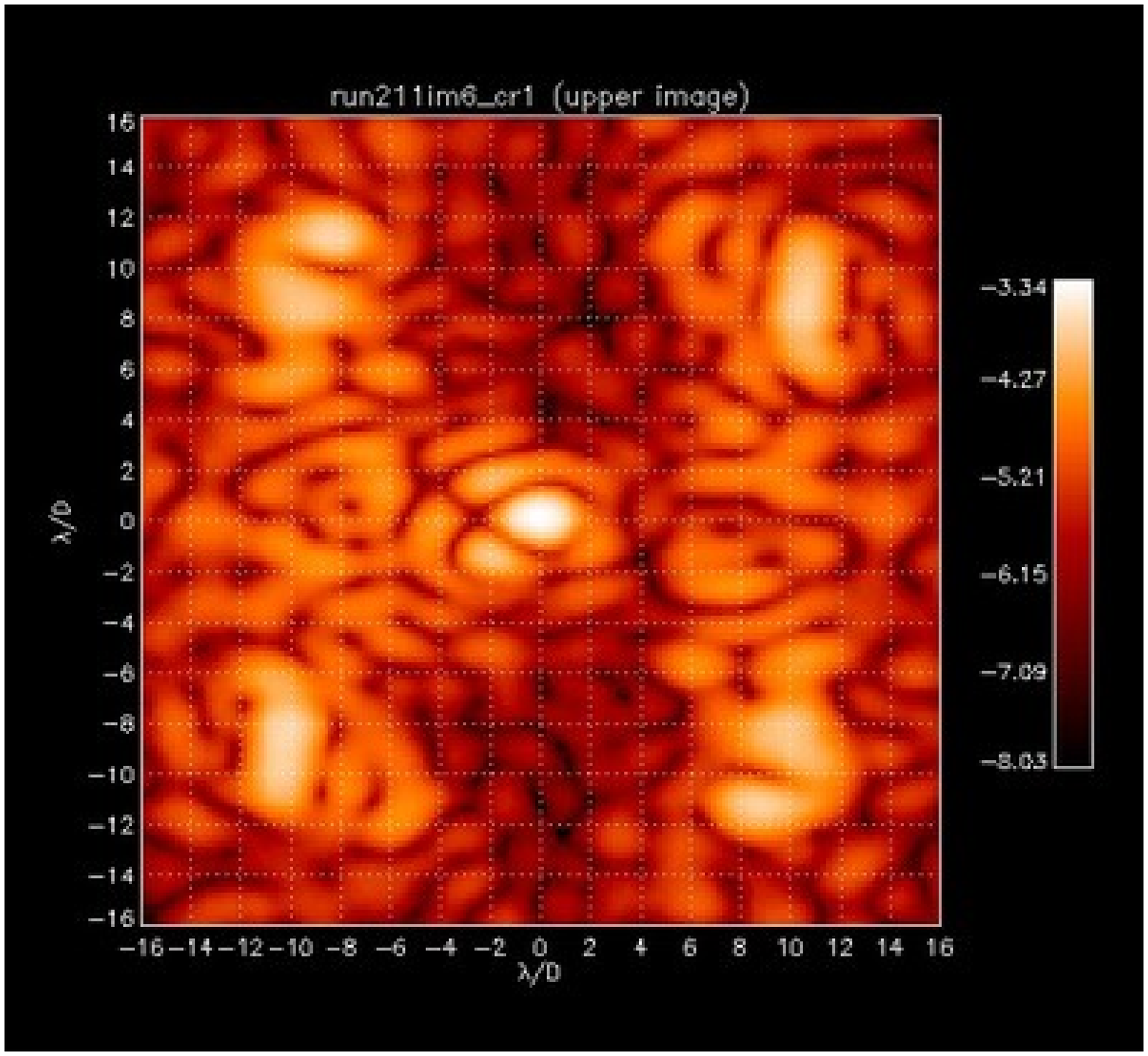} & \includegraphics[width=6cm]{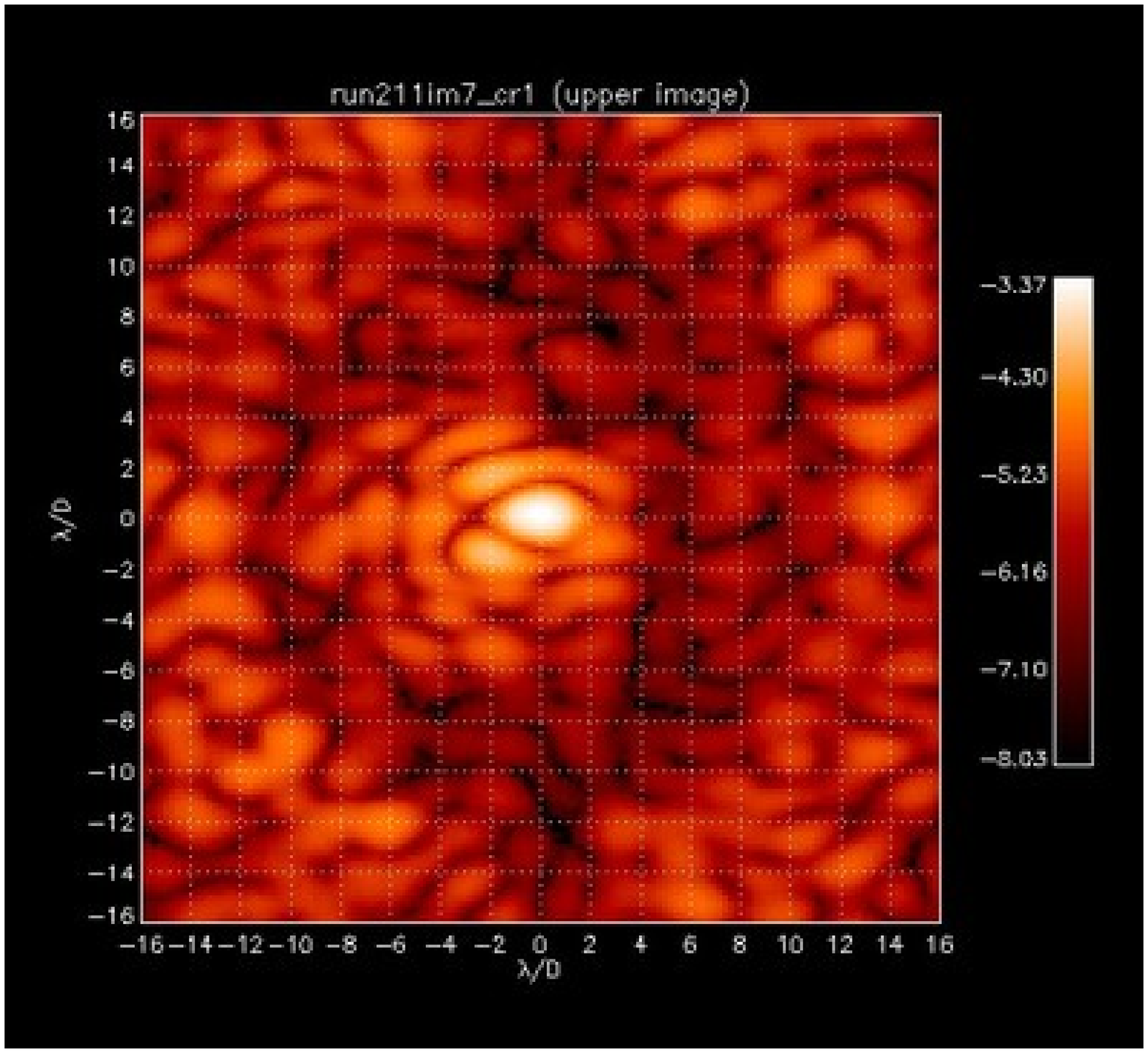}
\end{tabular}
\end{center}
\caption[example] { \label{fig:exp_im} 
Same as Fig.~\ref{fig:sim_im} but for a real experiment. The white rectangle ($[4\frac{\lambda}{D},12\frac{\lambda}{D}] \times [-12\frac{\lambda}{D},12\frac{\lambda}{D}]$) delimits the region where the algorithm is trying to create a dark hole. The initial image already contains an annular DH that was created by a different method.}
\end{figure}

In this section, we show monochromatic experiments at $\lambda = 785\:\mathrm{nm}$ with the High Contrast Imaging Testbed\cite{Trauger04} at JPL (Figs.~\ref{fig:exp_dm}--\ref{fig:exp_im}). For these, we have used the field nulling equation (Eq.~\ref{eq:field_nulling}) solved with an inverse FFT. In the initial image, the median and the mean in the DH are 73 and 146~ADU, respectively, corresponding to contrasts of $1.4 \times 10^{-6}$ and $6.8 \times 10^{-7}$, respectively. In the last image, the median and the mean in the DH are 72 and 139~ADU, respectively, corresponding to contrasts of $1.3 \times 10^{-6}$ and $6.7 \times 10^{-7}$, respectively. Unfortunately, the contrast remains almost unchanged in this first attempt. Nevertheless, it makes sense at this stage to try to iterate the measurement and correction process because some modeling approximations can be regarded as phase and amplitude errors. Indeed, after five iterations, the median and the mean in the DH are 51 and 112~ADU, respectively, corresponding to contrasts of $1.0 \times 10^{-6}$ and $4.8 \times 10^{-7}$, respectively. Simulations with a realistic noise level show that more iterations could have led to a median DH of $\sim 5$ ADU, about ten times better than what was obtained.

\section{CONCLUSION}
\label{sec:conclusion} 
Although our first experimental results are modest, they prove that our approach is sound and should be pursued. Future work will include an experimental demonstration of the energy minimization approach in monochromatic light, as well as a generalization of the theory to polychromatic light with corresponding experiments. Our wave front sensing and control theory is general and can be applied to other types of coronagraph, as demonstrated in Ref.~\citenum{Giveon06} for shaped-pupil coronagraphs.

Note that these experiments are by no means representative of the best result achieved to date with the HCIT. For these, we refer the reader to Ref.~\citenum{Trauger04} and to the paper by Trauger et al. in these proceedings.

\acknowledgments     
 
This work was performed in part under contract 1256791 with the Jet Propulsion Laboratory (JPL), funded by NASA through the Michelson Fellowship Program. JPL is managed for NASA by the California Institute of Technology. This research has made use of NASA's Astrophysics Data System.  


\bibliography{references}

\begin{thebibliography}{1}

\bibitem{Malbet95}
F.~{Malbet}, J.~W. {Yu}, and M.~{Shao}, ``{High-Dynamic-Range Imaging Using a
  Deformable Mirror for Space Coronography},'' {\em Pub. Astron. Soc.
  Pac.}~{\bf 107}, pp.~386--+, Apr.~1995.

\bibitem{Coulter04}
D.~R. {Coulter}, ``{NASA's Terrestrial Planet Finder missions},'' in {\em
  Microwave and Terahertz Photonics. Edited by Stohr, Andreas; Jager, Dieter;
  Iezekiel, Stavros. Proceedings of the SPIE, Volume 5487, pp. 1207-1215
  (2004).},  J.~C. {Mather}, ed., pp.~1207--1215, Oct.~2004.

\bibitem{Borde06}
P.~J. {Bord{\'e}} and W.~A. {Traub}, ``{High-Contrast Imaging from Space:
  Speckle Nulling in a Low-Aberration Regime},'' {\em Astrophys. J.}~{\bf 638},
  pp.~488--498, Feb.~2006.

\bibitem{Trauger03}
J.~T. {Trauger}, D.~{Moody}, B.~{Gordon}, Y.~{G{\"u}rsel}, M.~A. {Ealey}, and
  R.~B. {Bagwell}, ``{Performance of a precision high-density deformable mirror
  for extremely high contrast imaging astronomy from space},'' in {\em Future
  EUV/UV and Visible Space Astrophysics Missions and Instrumentation. Edited by
  J. Chris Blades, Oswald H. W. Siegmund. Proceedings of the SPIE, Volume 4854,
  pp. 1-8 (2003).},  J.~C. {Blades} and O.~H.~W. {Siegmund}, eds., pp.~1--8,
  Feb.~2003.

\bibitem{Press97}
W.~H. Press, S.~A. Teukolsky, W.~T. Vetterling, and B.~P. Flannery, {\em
  {Numerical Recipes in C, The Art of Scientific Computing}}, Cambridge
  University Press, second~ed., 1997.

\bibitem{Kuchner02}
M.~J. {Kuchner} and W.~A. {Traub}, ``{A Coronagraph with a Band-limited Mask
  for Finding Terrestrial Planets},'' {\em Astrophys. J.}~{\bf 570},
  pp.~900--908, May~2002.

\bibitem{Trauger04}
J.~T. {Trauger}, C.~{Burrows}, B.~{Gordon}, J.~J. {Green}, A.~E. {Lowman},
  D.~{Moody}, A.~F. {Niessner}, F.~{Shi}, and D.~{Wilson}, ``{Coronagraph
  contrast demonstrations with the high-contrast imaging testbed},'' in {\em
  Microwave and Terahertz Photonics. Edited by Stohr, Andreas; Jager, Dieter;
  Iezekiel, Stavros. Proceedings of the SPIE, Volume 5487, pp. 1330-1336
  (2004).},  J.~C. {Mather}, ed., pp.~1330--1336, Oct.~2004.

\bibitem{Giveon06}
A.~{Give'On}, N.~J. {Kasdin}, and R.~J. {Vanderbei}, ``{Closed-loop Wavefront
  Correction for High Contrast Imaging: The ``peak-a-boo'' Algorithm},'' in
  {\em Direct Imaging of Exoplanets: Science \& Techniques. Proceedings of the
  IAU Colloquium \#200, Edited by C. Aime and F. Vakili. Cambridge, UK:
  Cambridge University Press, 2006., pp.541-546},  C.~{Aime} and F.~{Vakili},
  eds., pp.~541--546, 2006.

\end{thebibliography}
\bibliographystyle{spiebib}

\end{document}